\documentclass[aps,prc,twocolumn,superscriptaddress,showpacs,nofootinbib]{revtex4-1}
\usepackage{graphicx,amsmath,amssymb,bm,color}

\newcommand{\mev}{\,\mathrm{MeV}}

\newcommand{\fm}{\,\mathrm{fm}}

\newcommand{\beq}{\begin{equation}}
\newcommand{\eeq}{\end{equation}}

\newcommand{\fet}[1]{\mbox{\boldmath $#1$}}

\allowdisplaybreaks

\begin{document}

\title{Quantum Monte Carlo calculations of neutron matter with chiral three-body forces}

\author{I.\ Tews}
\email[E-mail:~]{tews@theorie.ikp.physik.tu-darmstadt.de}
\affiliation{Institut f\"ur Kernphysik,
Technische Universit\"at Darmstadt, 64289 Darmstadt, Germany}
\affiliation{ExtreMe Matter Institute EMMI,
GSI Helmholtzzentrum f\"ur Schwerionenforschung GmbH, 64291 Darmstadt, Germany}
\author{S.\ Gandolfi}
\email[E-mail:~]{stefano@lanl.gov}
\affiliation{Theoretical Division, Los Alamos National Laboratory,
Los Alamos, NM 87545, USA}
\author{A.\ Gezerlis}
\email[E-mail:~]{gezerlis@uoguelph.ca}
\affiliation{Department of Physics,
University of Guelph, Guelph, Ontario N1G 2W1, Canada}
\author{A.\ Schwenk}
\email[E-mail:~]{schwenk@physik.tu-darmstadt.de}
\affiliation{Institut f\"ur Kernphysik,
Technische Universit\"at Darmstadt, 64289 Darmstadt, Germany}
\affiliation{ExtreMe Matter Institute EMMI,
GSI Helmholtzzentrum f\"ur Schwerionenforschung GmbH, 64291 Darmstadt, Germany}

\begin{abstract}
Chiral effective field theory (EFT) enables a systematic description
of low-energy hadronic interactions with controlled theoretical
uncertainties. For strongly interacting systems, quantum Monte Carlo
(QMC) methods provide some of the most accurate solutions, but they
require as input local potentials. We have recently constructed local
chiral nucleon-nucleon (NN) interactions up to next-to-next-to-leading
order (N$^2$LO). Chiral EFT naturally predicts consistent many-body
forces. In this paper, we consider the leading chiral three-nucleon
(3N) interactions in local form. These are included in auxiliary field
diffusion Monte Carlo (AFDMC) simulations. We present results for
the equation of state of neutron matter and for the energies and radii
of neutron drops. In particular, we study the regulator dependence at
the Hartree-Fock level and in AFDMC and find that present local
regulators lead to less repulsion from 3N forces compared to the usual
nonlocal regulators.
\end{abstract}


\maketitle

\section{Introduction}

Chiral effective field theory (EFT) provides a systematic expansion for nuclear 
forces based on
the symmetries of QCD~\cite{Epelbaum:2009a,Entem:2011,Hammer:2013}. At
a given order in the power counting, nuclear forces include
contributions from pion exchanges and from shorter-range interactions.
Chiral EFT enables calculations with controlled theoretical
uncertainties, a consistent description of electroweak interactions,
and the matching to lattice QCD. In addition to nucleon-nucleon (NN) interactions,
which are the dominant contribution to nuclear forces, chiral EFT
naturally predicts consistent many-body interactions, where
the leading three-nucleon (3N) forces enter at next-to-next-to-leading order 
(N$^2$LO)~\cite{3Nforces1,3Nforces2}. It
has been shown that 3N forces are key for the properties of neutron
and nuclear matter~\cite{Hebeler:2010a,Hebeler:2010xb,Gandolfi:2012,%
Holt:2013fwa,Tews:2013,Hagen:2014,Wellenhofer:2014hya,Carbone:2014,%
Sammarucca:2014}. A better understanding of 3N forces is a major
frontier in nuclear physics.

In addition to systematic chiral EFT interactions, reliable many-body
methods are needed. For strongly interacting systems, quantum Monte Carlo 
(QMC) methods provide some of the most accurate
solutions~\cite{Ceperley:1995,Carlson:2012}. These include Green's
function Monte Carlo
(GFMC)~\cite{Pudliner1997,Pieper:2001,Nollett2007} and auxiliary field
diffusion Monte Carlo (AFDMC)~\cite{Schmidt1999} methods; for a recent 
review see Ref.~\cite{Carlson:2014}. In continuum QMC calculations, the 
central object is the many-body propagator, which is of the form
\begin{equation}
G({\bf R},{\bf R}^{\prime}; \delta \tau) =
\langle{\bf R}|e^{-\delta \tau \widehat{H}}|
{\bf R}^{\prime}\rangle \,.
\end{equation}
Here, ${\bf R} = ({\bf r}_{1},s_1,{\bf r}_{2},s_2,\ldots, {\bf
r}_{N},s_N)$ is the configuration vector of all $N$ particles,
including the single-particle coordinates and spins $\textbf{r}_i, s_i$ (and other quantum numbers), $\delta
\tau$ is a step in the imaginary-time evolution, and $\widehat{H}$ is
the Hamiltonian. 

In nuclear GFMC calculations all possible spin-isospin nucleon
states need to be explicitly accounted for, which makes this method
unsuitable for accurate neutron matter studies due to an unfavorable
scaling behavior. In contrast to GFMC, AFDMC rewrites the propagator
by applying a Hubbard-Stratonovich transformation using
auxiliary fields, which changes the scaling behavior favorably at the
cost of additional integrations over auxiliary fields. We thus make
use of the AFDMC method to study neutron matter.

The trial wave function $\psi_T$ in AFDMC is usually chosen to be of the form
\begin{equation}
\psi_T(\textbf{R}) = \left[\prod_{i<j} f_J(r_{ij}) \right] 
\Phi_A(\textbf{R}) \,,
\end{equation}
where inter-particle correlations are included through the Jastrow
factor $f_J(r_{ij})$ and $\Phi_A$ is the noninteracting ground state
given by a Slater determinant
\begin{equation}
\Phi_A(\textbf{R}) = \mathcal{A} \left[\prod_i 
\phi_{\alpha_i}(\textbf{r}_i,s_i) \right] \,,
\end{equation}
where $\alpha_i$ labels single-particle states (plane waves for 
neutron matter and Hartree-Fock orbitals for neutron drops 
\cite{Maris:2013rgq}).

For the evaluation of the propagator, it is necessary
to be able to separate all momentum dependences up to quadratic
terms.  This can be done for local
interactions, where the propagator for the momentum-dependent part is
a Gaussian integral that can be evaluated analytically, while the
interaction part can be easily obtained from the configuration vector
(for more details see Ref.~\cite{Carlson:2014}). Chiral EFT
interactions are naturally formulated in momentum space and usually
contain several sources of nonlocality.

Recently, local chiral NN potentials have been constructed up to
N$^2$LO and have been used to calculate the energy of neutron matter
and light nuclei using continuum QMC
methods~\cite{Gezerlis:2013ipa,Gezerlis:2014ipa,Lynn2014}. Following
the same strategy, a minimally nonlocal NN potential was developed in
Ref.~\cite{Piarulli} with explicit $\Delta$ degrees of freedom. Monte
Carlo methods have also been used to study neutron matter based on
lattice techniques and other momentum-space
QMC approaches~\cite{nuclattice,Roggero,Bulgac}.

For a complete calculation at N$^2$LO, it is necessary to include 3N
forces. In Sec.~\ref{sec:2}, we present local chiral 3N forces at
N$^2$LO, which are consistent with the local NN interactions of
Refs.~\cite{Gezerlis:2013ipa,Gezerlis:2014ipa}. The general
expressions for the local 3N forces are given in the
Appendix. We study in detail the regulator dependence of
the leading two-pion-exchange 3N energy contributions at the
Hartree-Fock level in Sec.~\ref{sec:3} and in AFDMC in
Sec.~\ref{sec:4}.  This shows that present local regulators lead to
less repulsion from 3N forces compared to using the usual nonlocal
regulators. We present results for the equation of state of neutron
matter in Sec.~\ref{sec:4} and for the energies and radii of
neutron drops in Sec.~\ref{sec:5}. Finally, we summarize and give
an outlook.

\section{Chiral 3N forces in coordinate space}
\label{sec:2}

\begin{figure}[t]
\centering
\includegraphics[width=0.45\textwidth]{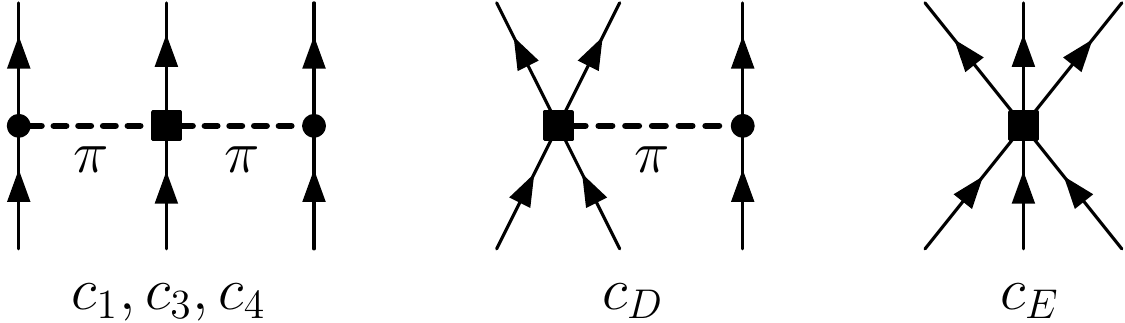}
\caption{Contributions to 3N forces at N$^2$LO. These include
a two-pion-exchange part given by the couplings $c_1, c_3,$
and $c_4$, a one-pion-exchange--contact interaction given by
$c_D$, and a 3N contact interaction given by $c_E$.\label{fig:N$^2$LO_3N}}
\end{figure}

In chiral EFT, the leading 3N forces at N$^2$LO have three
contributions: a two-pion-exchange part given by the couplings $c_1,
c_3,$ and $c_4$, a one-pion-exchange--contact interaction given by
$c_D$, and a 3N contact interaction given by $c_E$
\cite{3Nforces1,3Nforces2}. We show the N$^2$LO 3N contributions
diagrammatically in Fig.~\ref{fig:N$^2$LO_3N}. The $c_i$ couplings are
determined by pion-nucleon or NN scattering, while $c_D$ and $c_E$
have to be fit to properties of $A > 2$ systems.

Because we want to include 3N forces in AFDMC calculations in
coordinate space, we have to use local coordinate-space expressions of
the N$^2$LO 3N forces, as was similarly done in
Refs.~\cite{Lovato:2012,Navratil:2007}. To achieve this, we first
Fourier transform the momentum-space expressions of the N$^2$LO 3N
forces. We begin with the 3N contact interaction $V_E$. In momentum
space, this contribution vanishes in neutron matter due to the Pauli
principle, when a regulator that is symmetric in the particle labels
is used; see Ref.~\cite{Hebeler:2010a}. Because a local regulator does
not fulfill this requirement, the $V_E$ term will contribute. In this
case, the regulator induces a finite range that mixes 3N partial waves.
After Fourier transformation, we find in neutron matter (with
${\fet \tau}_i \cdot {\fet \tau}_k = 1$)
\begin{align}
\quad V_E^{ijk} =\frac{c_E}{2 f_{\pi}^4 \Lambda_{\chi}} \sum_{\pi(ijk)} \,
\delta(\textbf{r}_{ij}) \delta(\textbf{r}_{kj}) \,,
\end{align}
where we sum over all permutations $\pi(ijk)$ of the three particles
$i, j,$ and $k$, $\textbf{r}_{ij} = \textbf{r}_i - \textbf{r}_j$,
$f_{\pi}=92.4 \mev$ is the pion decay constant, and we use
$\Lambda_{\chi}=700 \mev$. The expressions for general isospin and
details on the Fourier transformation are provided in the Appendix.

As for the $V_E$ term, the one-pion-exchange--contact interaction
$V_D$ vanishes in momentum space for neutron matter due to the
spin-isospin structure, if a symmetric regulator is
used~\cite{Hebeler:2010a}. In coordinate space, the $V_D$ term also
contributes and after Fourier transformation we have two parts
(see the Appendix):
\begin{align}
V_{D}^{ijk} &= \frac{g_A}{24f_{\pi}^4}\frac{c_D}{\Lambda_{\chi}}
\sum_{\pi(ijk)} \biggl[ \frac{m_{\pi}^2}{4 \pi} \, \delta(\textbf{r}_{ij}) X_{ik}(\textbf{r}_{kj}) \nonumber \\
&\quad - {\fet \sigma}_i \cdot {\fet \sigma}_k \, \delta(\textbf{r}_{ij}) \delta(\textbf{r}_{kj}) \biggr] \,,
\end{align}
where $g_A=1.267$ is the axial coupling, $m_{\pi}=138.03 \mev$
is the averaged pion mass, and the function $X_{ik}(\textbf{r})$
is given by
\begin{equation}
X_{ik}(\textbf{r}) = \bigl[ S_{ik}(\textbf{r}) \, T(r)
+ {\fet \sigma}_i \cdot {\fet \sigma}_k \bigr] Y(r) \,,
\label{eq:X}
\end{equation}
with the tensor operator $S_{ik}({\bf r}) = 3 \fet {\fet \sigma}_i
\cdot \widehat{\bf r}{\fet \sigma}_k \cdot \widehat{\bf r} - {\fet
\sigma}_i \cdot {\fet \sigma}_k$, the function $T(r)=1+3/(m_{\pi}
r)+3/(m_{\pi} r)^2$, and the Yukawa function $Y(r)=e^{-m_{\pi}
r}/r$. Here, $\widehat{\textbf{r}}$ is the unit vector in the
direction of $\textbf{r}$ and $r$ is the magnitude. There are two
contributions from $V_D$, because one-pion exchange contains a
long-range part as well as a delta-function part. The latter needs to
be included to maintain the Goldstone boson nature of the pion.

We emphasize that there is an ambiguity in performing the Fourier
transformation for the $V_D$ term, depending on the choice of the
initial spin-isospin structure. This leads either to terms
$\delta(\textbf{r}_{ij}) X_{ik}(\textbf{r}_{kj})$ or
$\delta(\textbf{r}_{ij}) X_{kj}(\textbf{r}_{kj})$ with different spin
indices in the $X$ function. The two expressions are the same due to
the $\delta$ function, but lead to different results after
regularization. Therefore, the differences from choosing different
structures are a regulator effect and should be of higher order, as
mentioned in Ref.~\cite{Navratil:2007}. These differences will vanish
in the limit of infinite momentum cutoff. Because we will not include
the $V_D$ term in our calculations in this paper, this effect will not
influence the results here. The different $V_D$ terms are studied
in Ref.~\cite{Lynn2015}.

We now turn to the two-pion-exchange contributions $V_C$ in neutron
matter (see the Appendix). For the part proportional to
$c_1$, we find
\begin{align}
V_{C,c_1}^{ijk} &= \frac{c_1 m_{\pi}^4 g_A^2}{2 f_{\pi}^4 (4 \pi)^2}
\sum_{\pi(ijk)} {\fet \sigma}_i \cdot \hat{\textbf{r}}_{ij} \, {\fet \sigma}_k \cdot \hat{\textbf{r}}_{kj} \, \nonumber \\
&\quad \times  U(r_{ij}) Y(r_{ij}) U(r_{kj}) Y(r_{kj}) \,,
\label{eq:contributionc1}
\end{align}
with the function $U(r)=1+1/(m_{\pi} r)$. This contribution is similar
to the long-range (LR) $S$-wave part of the Illinois 3N forces; see
Ref.~\cite{Pieper2001}.

The part proportional to $c_3$ is given by
\begin{align}
V_{C,c_3}^{ijk} &= \frac{c_3 g_A^2}{36 f_{\pi}^4} \sum_{\pi(ijk)} \nonumber \\
&\quad \times
\biggl[\frac{m_{\pi}^4}{(4 \pi)^2} \, X_{ij}(\textbf{r}_{ij}) X_{kj}(\textbf{r}_{kj}) - \frac{m_{\pi}^2}{4 \pi} \, X_{ik}(\textbf{r}_{ij}) \delta(\textbf{r}_{kj}) \nonumber \\[1mm]
&\quad - \frac{m_{\pi}^2}{4 \pi} \, X_{ik}(\textbf{r}_{kj}) \delta(\textbf{r}_{ij}) + {\fet \sigma}_i \cdot {\fet \sigma}_k  \delta(\textbf{r}_{ij}) \, \delta(\textbf{r}_{kj}) \biggr] \,.
\label{eq:contributionc3}
\end{align}
Thus, in coordinate space, for the $c_3$ part of the $V_C$ term, there
are four contributions due to a long-range and short-range part in
each pion exchange. The first term $\sim X_{ij}(\textbf{r}_{ij})
X_{kj}(\textbf{r}_{kj})$ is a long-range two-pion-exchange
contribution similar to the anticommutator part of the $P$-wave
two-pion-exchange interaction of Ref.~\cite{Pieper2001}.  In addition,
there is also a short-range (SR) term $\sim \delta(\textbf{r}_{ij})
\delta(\textbf{r}_{kj})$ which is similar to $V_E$ but spin-dependent,
and two intermediate-range (IR) terms $\sim X_{ik}(\textbf{r}_{ij})
\delta(\textbf{r}_{kj}) + X_{ik}(\textbf{r}_{kj})
\delta(\textbf{r}_{ij})$ similar to $V_D$. Note that although the
spin-isospin structure is similar to the Urbana IX force and in
general to the two-pion-exchange part of the Illinois forces, the
spatial functions are quite different. Finally, the coordinate-space
expression for the $c_4$ part of $V_C$ is given in the
Appendix. This does not contribute in neutron matter for
general regulators due to the isospin structure.

\begin{figure}[t]
\centering
\includegraphics[height=0.48\textwidth]{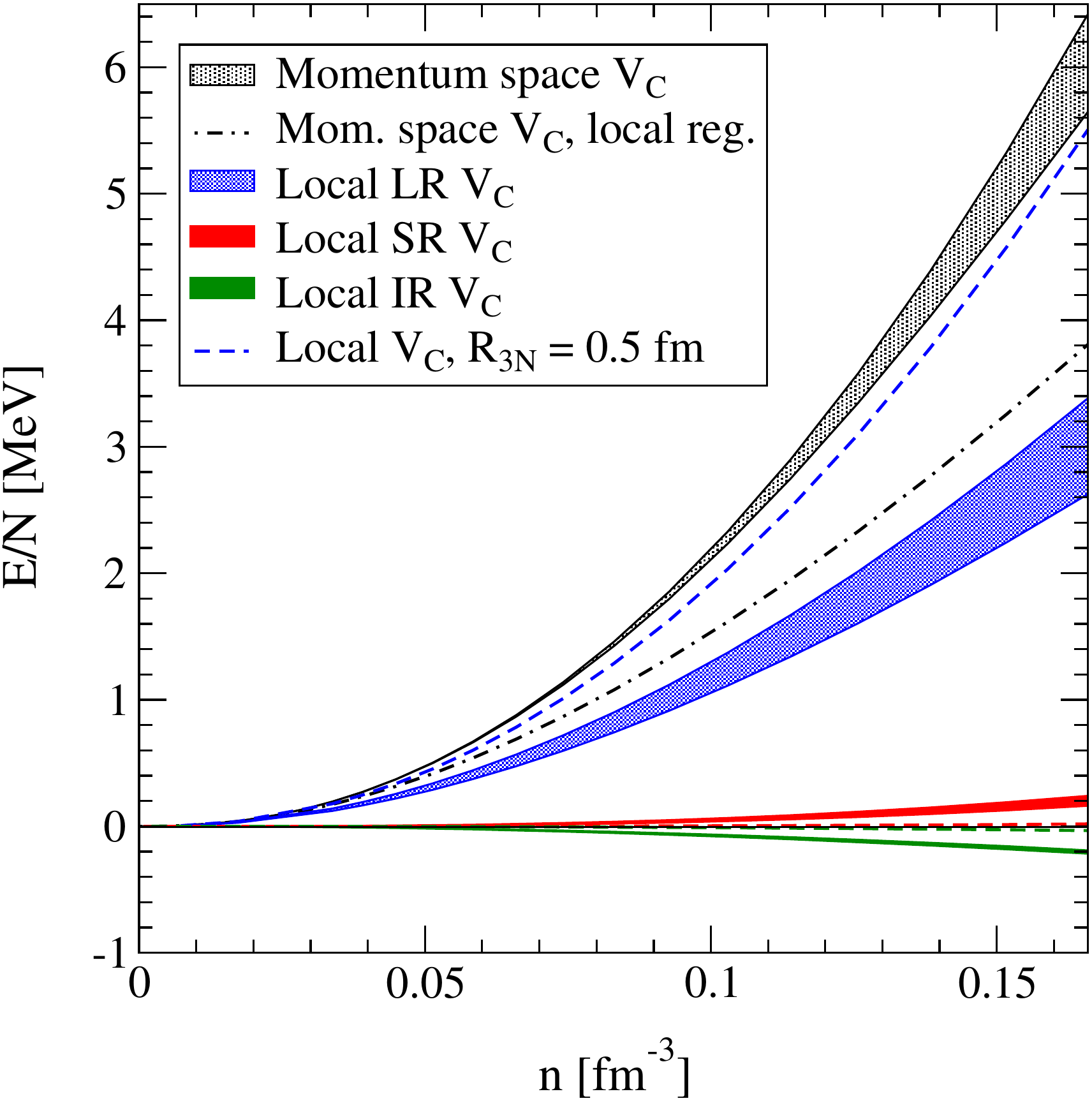}
\caption{Contributions to the neutron-matter energy per particle $E/N$ as a
function of density $n$ at the Hartree-Fock level. The black band
shows the energy obtained using a nonlocal regulator, as in
Ref.~\cite{Kruger:2013}, with a 3N cutoff $400-500 \mev$. The blue band
corresponds to the LR part of the two-pion-exchange interaction $V_C$
with the local regulator used here, the red band to the SR part of
$V_C$, and the green band to the IR of $V_C$. For these bands, the
cutoff in the local regulator is varied with
$R_\text{3N} = 1.0-1.2 \fm$. The dashed-dotted line corresponds to
the results for $V_C$ using the local momentum-space regulator of
Ref.~\cite{Navratil:2007} with a cutoff $\Lambda_\text{3N} =500
\mev$. This shows that these local 3N forces provide less repulsion at
the Hartree-Fock level than with nonlocal regulators. The dashed
lines show the results for $V_C$ with the local regulator and small
$R_\text{3N} = 0.5 \fm$.\label{fig:HF_reg}}
\end{figure}

\begin{figure}[t]
\centering
\includegraphics[height=0.48\textwidth]{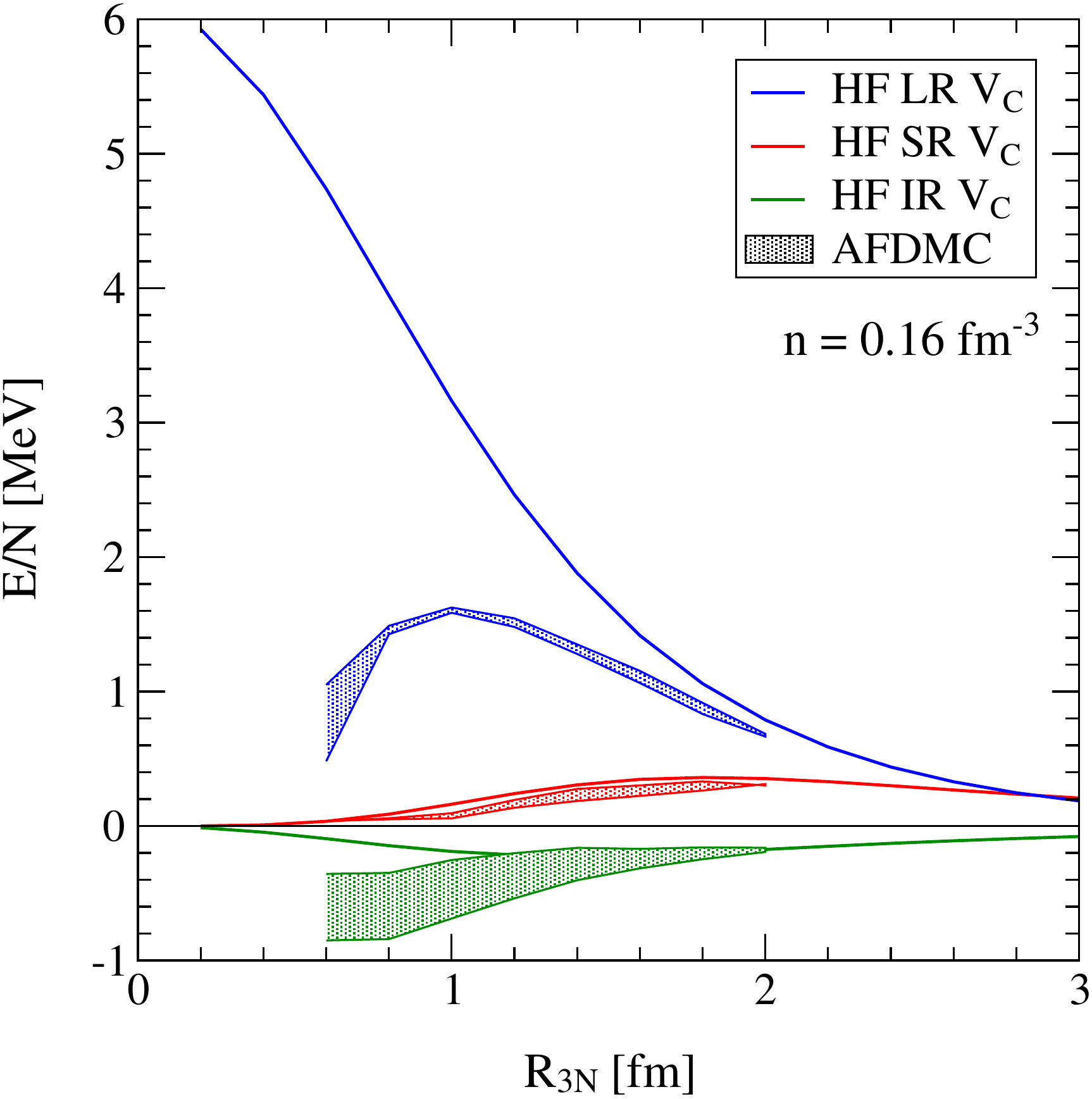}
\caption{Contributions to the energy per particle $E/N$ at saturation density
as a function of the cutoff $R_\text{3N}$. The lines show the LR, SR,
and IR parts of the two-pion-exchange interaction $V_C$ with the local
regulator used here, calculated at the Hartree-Fock level.  The bands
are the contributions of the corresponding 3N parts to the AFDMC
energies for a variation of the NN cutoff $R_0=1.0-1.2 \fm$; see also
Fig.~\ref{fig:cutoff}.\label{fig:HF_reg_comp}}
\end{figure}

For a many-body system, the total 3N interactions are then given by
$V_{\rm 3N} = \sum_{i<j<k} V_{\rm 3N}^{ijk}$, with $i,j,k = 1, \ldots, A$.
Moreover, in the AFDMC calculation $V_{\rm 3N}^{ijk}$ is rewritten as a
sum over cyclic permutations only.

In order to regularize the local 3N forces consistently with the NN
forces of Refs.~\cite{Gezerlis:2013ipa,Gezerlis:2014ipa}, we replace
the $\delta$ functions by smeared-out delta functions of the form
\begin{equation}
\delta ({\bf r}) \; \to \; \delta_{R_{3\text{N}}} ({\bf r})
= \frac{1}{\pi \Gamma\bigl(3/4\bigr) R_{3\text{N}}^3} e^{-(r/R_{3\text{N}})^4} \,,
\end{equation}
where $R_{\text{3N}}$ is the three-body cutoff. For the long-range
pion-exchange contributions, we multiply the Yukawa functions with the
long-range regulator $f_{\text{long}}$ of
Refs.~\cite{Gezerlis:2013ipa,Gezerlis:2014ipa}, given by
\begin{equation}
Y(r) \; \to \; Y(r) \Bigl( 1 - e^{-(r/R_{3\text{N}})^4} \Bigr) \,.
\end{equation}
To be consistent with the NN cutoff $R_0=1.0 - 1.2 \fm$, we will also
vary the 3N cutoff in this range, $R_{\text{3N}}=1.0 - 1.2 \fm$. We have
checked that the IR and SR parts of $V_C$ as well as the $V_E$ and $V_D$
contributions in neutron matter vanish for $R_{\text{3N}} \to 0$ (for
infinite momentum cutoffs), and are therefore regulator effects.

In the following, we include all terms of $V_C$ (the $c_1$ and $c_3$
parts for neutron matter), with the $c_i$ couplings having the same values
as in the local NN interactions~\cite{Gezerlis:2013ipa,Gezerlis:2014ipa}. Results
including the shorter-range contributions $V_D$ and $V_E$, which
require fits of $c_D$ and $c_E$, are studied in
Ref.~\cite{Lynn2015}.

\section{Hartree-Fock calculation for neutron matter}
\label{sec:3}

We calculate the 3N contributions from the $V_C$ part to neutron
matter first at the Hartree-Fock (HF) level. This includes all
interactions of Eqs.~(\ref{eq:contributionc1}) and
(\ref{eq:contributionc3}). Details on the HF calculation can be found
in Refs.~\cite{Kruger:2013}.

In Fig.~\ref{fig:HF_reg} we show the contributions to the
neutron-matter energy per particle $E/N$ as a function of
density~$n$. The blue band corresponds to the LR part of the
two-pion-exchange interaction $V_C$ with the local regulator used
here, the red band to the SR part of $V_C$, and the green band to the
IR of $V_C$. For these bands, the cutoff in the local regulator is
varied between $R_\text{3N} = 1.0-1.2 \fm$. The dashed lines show the
results for $V_C$ with the local regulator and $R_\text{3N} = 0.5
\fm$. In addition, the black band shows the energy obtained using a
nonlocal regulator, as in Ref.~\cite{Kruger:2013}, with a cutoff
$400-500 \mev$.

\begin{figure}[t]
\centering
\includegraphics[height=0.48\textwidth]{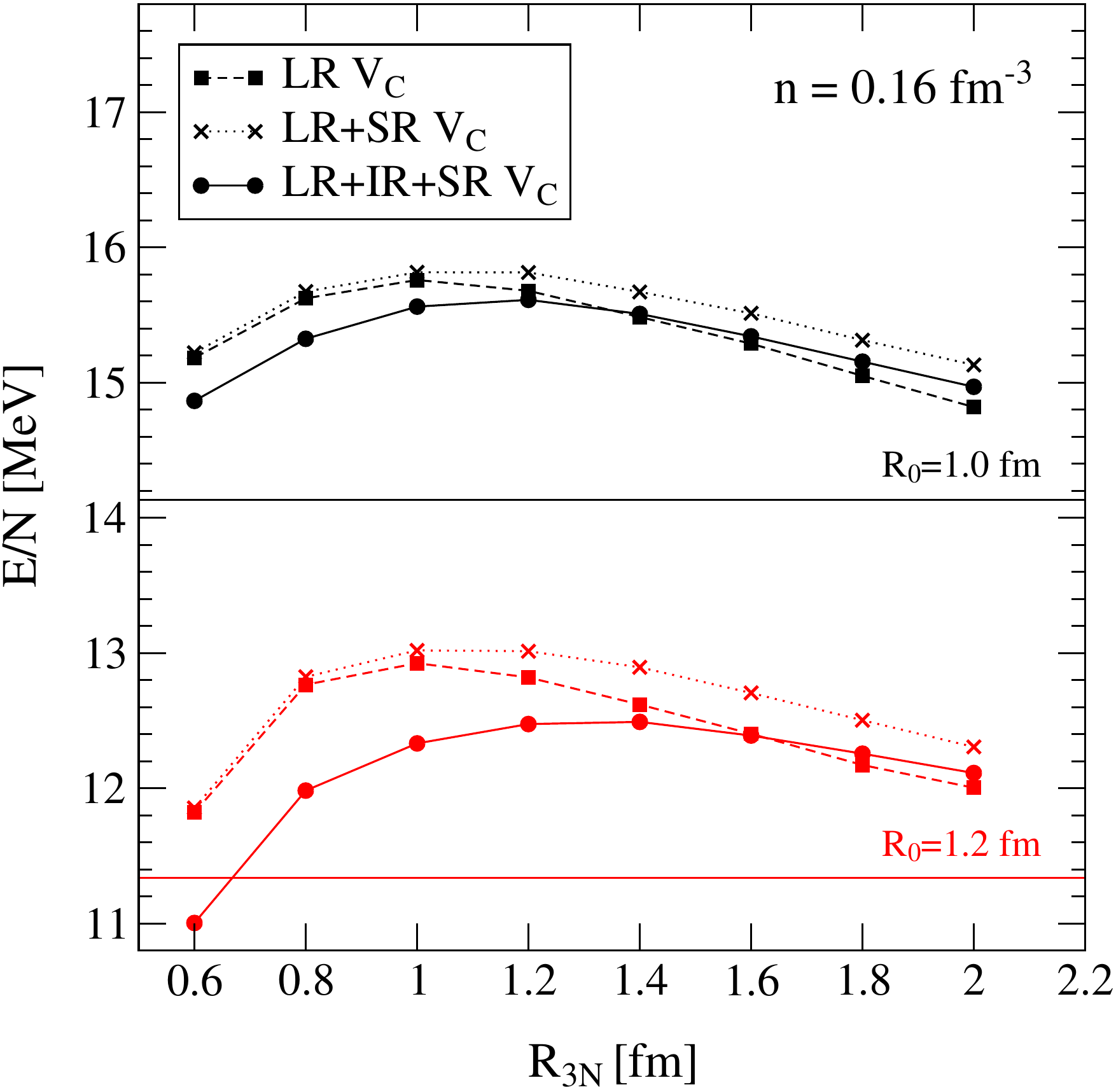}
\caption{Variation of the AFDMC energy per particle
at saturation density as a function of the 3N cutoff $R_\text{3N}$ for
an NN cutoff $1.0 \fm$ (black lines in the upper part) and $1.2 \fm$
(red lines in the lower part). The horizontal lines correspond to the
NN-only energy. The squares are for the $c_1$ and LR $c_3$ part of
$V_C$, the crosses include also the SR $c_3$ part of $V_C$, and the
circles include all parts of $V_C$.\label{fig:cutoff}}
\end{figure}

\begin{figure}[t]
\centering
\includegraphics[height=0.48\textwidth]{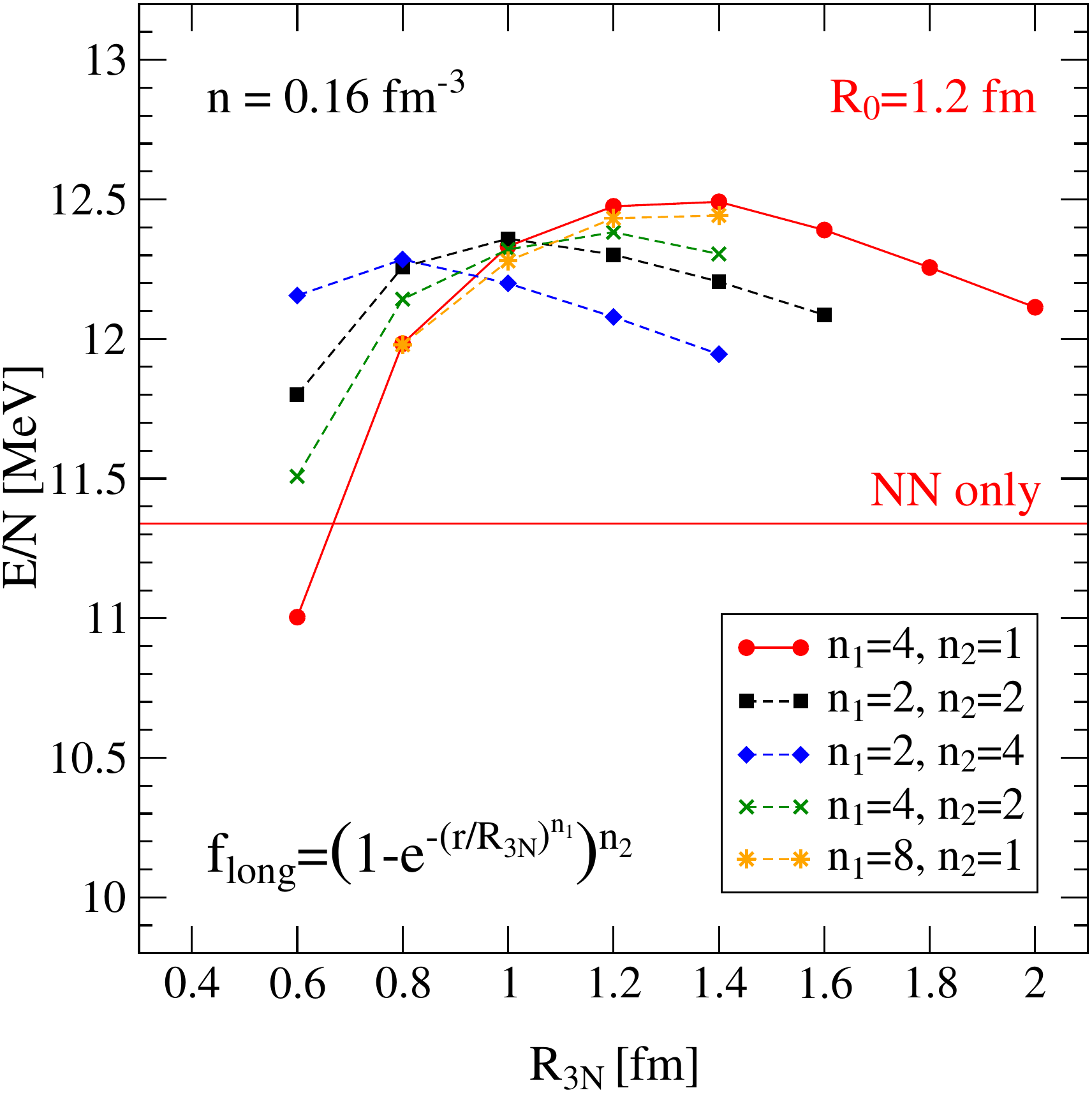}
\caption{Dependence of the AFDMC energy per particle
at saturation density as a function of the 3N cutoff $R_\text{3N}$
on different long-range regulators. Results are shown for an NN
cutoff $R_0=1.2 \fm$. The long-range regulator is given by
$\bigl[1-e^{-(r/R_{3\text{N}})^{n_1}}\bigr]^{n_2}$ with different
parameters $n_1$ and $n_2$.\label{fig:cutoff_reg_var}}
\end{figure}

The HF energy in neutron matter for the local $V_C$ are in total
$\approx 3 \mev$ at saturation density $n_0=0.16 \fm^{-3}$.  This is
only about half of the $V_C$ contribution using the nonlocal
regulator.  The shorter-range (IR and SR) contributions, which are
regulator effects, are small and with opposite sign. If we lower the
coordinate-space cutoff, $R_{3\text{N}} = 0.5 \fm$ (dashed lines), we
find that the IR and SR parts almost vanish, as expected, and that the
total HF energy is $5.5\mev$ for the local $V_C$, which agrees well
with the momentum-space result. We also note that the momentum-space
result is very close to the infinite-cutoff result at the HF
level. Thus, the smaller 3N energies for the local 3N forces are due
to the local regulators used.

To check this, we have performed a HF calculation of $V_C$ using the
local momentum-space regulator of Ref.~\cite{Navratil:2007} with a
cutoff of $\Lambda_{3\text{N}}=500 \mev$. This is given by the
dashed-dotted line in Fig.~\ref{fig:HF_reg}. At saturation density, we
find an energy per particle of $3.8 \mev$, which is comparable to the
result for the local 3N forces used here. This supports the above
conclusion that 3N forces with local regulators provide less repulsion
at the HF level compared to the usual nonlocal regulators. It may be
possible that the $V_D$ and $V_E$ parts, which contribute to neutron
matter for local regulators, make up part of these differences. This
is explored further in Ref.~\cite{Lynn2015}.

We show the $V_C$ contributions to the energy per particle $E/N$ at
saturation density as a function of the 3N cutoff $R_\text{3N}$ in
Fig.~\ref{fig:HF_reg_comp}. The lines show the LR, SR, and IR parts
for the local regulator used here, calculated at the HF level. For all
3N cutoffs, the SR and IR parts are small and of opposite sign, while
the major contribution of $V_C$ comes from the LR parts. The SR and IR
parts vanish for small coordinate-space (high momentum-space) cutoffs,
as expected. The LR part increases up to the infinite-cutoff result at
the HF level. In the cutoff range $R_{\text{3N}}=1.0-1.2 \fm $, the
total contribution of $V_C$ is $\approx 3 \mev$, and thus only about
half of infinite-cutoff result. This is what we also found in
Fig.~\ref{fig:HF_reg}. We emphasize that the cutoff dependence from
$400 - 500 \mev$ to infinte momentum-space cutoff is small for
nonlocal regulators and these densities.

\section{QMC calculation for neutron matter}
\label{sec:4}

\begin{figure}[t]
\centering
\includegraphics[height=0.48\textwidth]{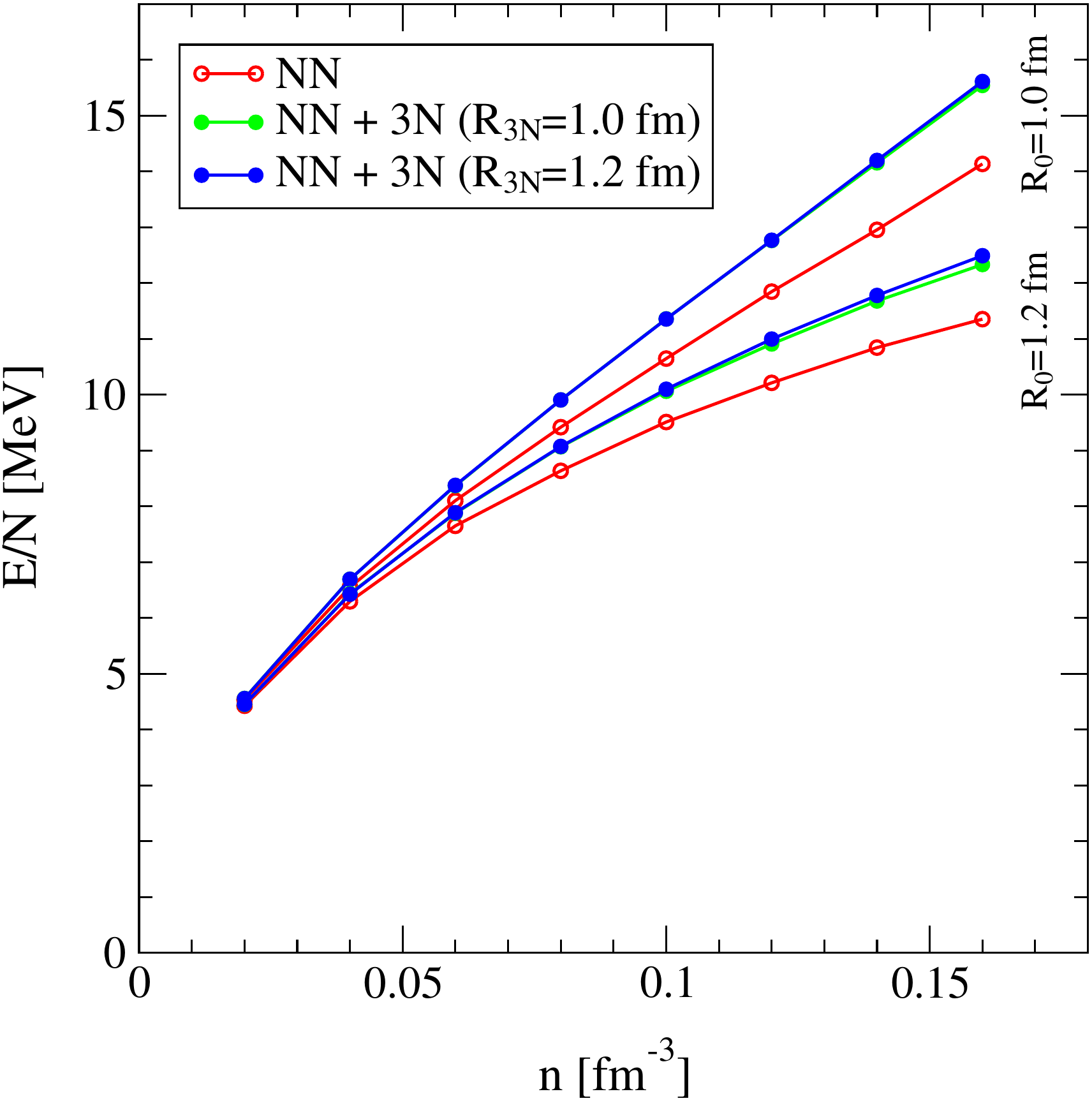}
\caption{Energy per particle as a function of density
for neutron matter at N$^2$LO, including NN forces and the 3N
$V_C$ interaction in AFDMC. Results are shown for an NN cutoff
$R_0=1.0-1.2 \fm$ and $R_{3\text{N}}$ in the same
range.\label{fig:n2loEOS}}
\end{figure}

\begin{figure}[t]
\centering
\includegraphics[height=0.48\textwidth]{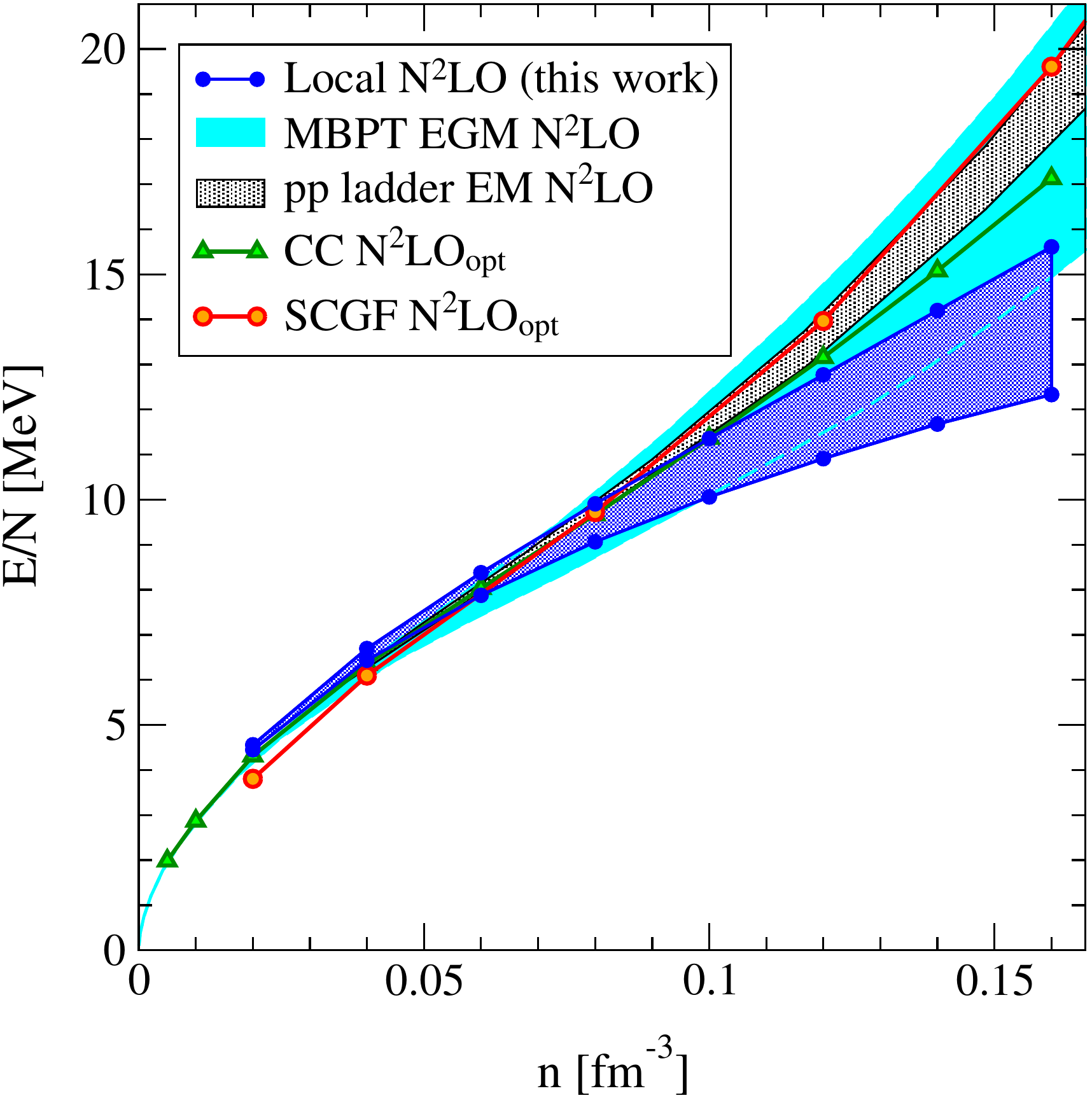}
\caption{Comparison of the neutron-matter energy at
N$^2$LO based on the local chiral NN+3N potentials in AFDMC (this
work) with the N$^2$LO calculation of Ref.~\cite{Tews:2013} based on
the EGM N$^2$LO potentials and using MBPT, with the particle-particle (pp) 
ladder results of
Ref.~\cite{Sammarucca:2014} based on the EM N$^2$LO potential, and
with results based on the N$^2$LO$_\text{opt}$ potential using
CC theory~\cite{Hagen:2014} and the SCGF method~\cite{Carbone:2014}.
\label{fig:n2lo_comparison}}
\end{figure}

Next, we investigate 3N forces in neutron matter using the AFDMC
method, similarly to
Refs.~\cite{Gezerlis:2013ipa,Gezerlis:2014ipa}. In
Fig.~\ref{fig:HF_reg_comp}, in addition to the HF results, we show the
contributions of the LR, SR, and IR parts of $V_C$ to the AFDMC
energy, where the bands are from varying the NN cutoff $R_0=1.0-1.2
\fm$. For the SR part, the AFDMC energies agree well with the HF
energies, so that HF is a good approximation for this
contribution. For the IR part, for large coordinate-space cutoffs, the
agreement between HF and AFDMC results is good but worsens for smaller
cutoffs. The uncertainty (from NN cutoff variation) grows and the
energy decreases significantly compared to the HF result. For the LR
parts, the AFDMC energies are about $70-80 \%$ of the HF energies for
higher cutoffs, which suggests that the LR N$^2$LO 3N contributions
beyond HF are important. When lowering the 3N cutoff, the energy
increases up to a plateau. Further lowering the cutoff, the system
collapses and the energy rapidly decreases. In addition, the
uncertainty grows.

To study these effects more clearly, we show the variation of the
total AFDMC energy per particle at saturation density as a function of
the 3N cutoff $R_\text{3N}$ for an NN cutoff $1.0 \fm$ (black lines in
the upper part) and $1.2 \fm$ (red lines in the lower part) in
Fig.~\ref{fig:cutoff}. The horizontal lines correspond to the NN-only
energies. The squares include only the LR $c_1$ and $c_3$ part of
$V_C$, the crosses include also the SR $c_3$ part of $V_C$, and the
circles include all parts of $V_C$. For the soft NN potential
($R_0=1.2 \fm$), we find the plateau of the AFDMC energy to be at
$R_{3\text{N}}=1.2-1.4 \fm$. If the cutoff is lowered, the energy
decreases and for $R_{3\text{N}}=0.6 \fm$ we find an attractive 3N
contribution.\footnote{This behavior is qualitatively similar to the
  overbinding given by the Illinois 3N forces in pure neutron
  systems~\cite{Sarsa:2003}.  It would be interesting to see if using
  a similar cutoff would avoid the overbinding of neutron matter using
  Illinois forces.} For the harder NN potential ($R_0=1.0 \fm$), the
plateau is found for smaller 3N cutoffs, $R_{3\text{N}}=1.0-1.2 \fm$.

In general, the plateau is reached when $R_{3\text{N}} \sim R_0$, and
the system collapses when $R_{3\text{N}}$ is significantly smaller
than $R_0$. This can be understood because harder NN potentials do not
favor particles to be close and smaller 3N cutoffs are needed to
overcome this repulsion. If we want to decrease $R_{3\text{N}}$ in our
calculations, we also need to decrease $R_0$. Therefore,
$R_{3\text{N}}$ has to be chosen consistently with $R_0$ which
justifies our cutoff range. Because $R_0 < 1.0 \fm$ is difficult for
local NN potentials~\cite{Gezerlis:2014ipa}, we also do not decrease
the 3N cutoff below that limit and use the range for $R_0$ and
$R_{3\text{N}}$ within $1.0-1.2 \fm$.

Although the collapse of the system for lower 3N cutoffs does not appear
in the HF calculation, it is not an artifact of the AFDMC method.
It is due to the function $X_{ij}(\textbf{r})$, which includes terms
$\sim 1/r^3$. If three particles are in a small volume, 
this becomes very attractive, unless it gets regularized with a large enough $R_{3\text{N}}$.
These cutoff values correspond to the position of the plateau. In a HF calculation, which only includes low-momentum states, this collapse will not appear.

\begin{figure*}[t]
\centering
\includegraphics[height=0.48\textwidth]{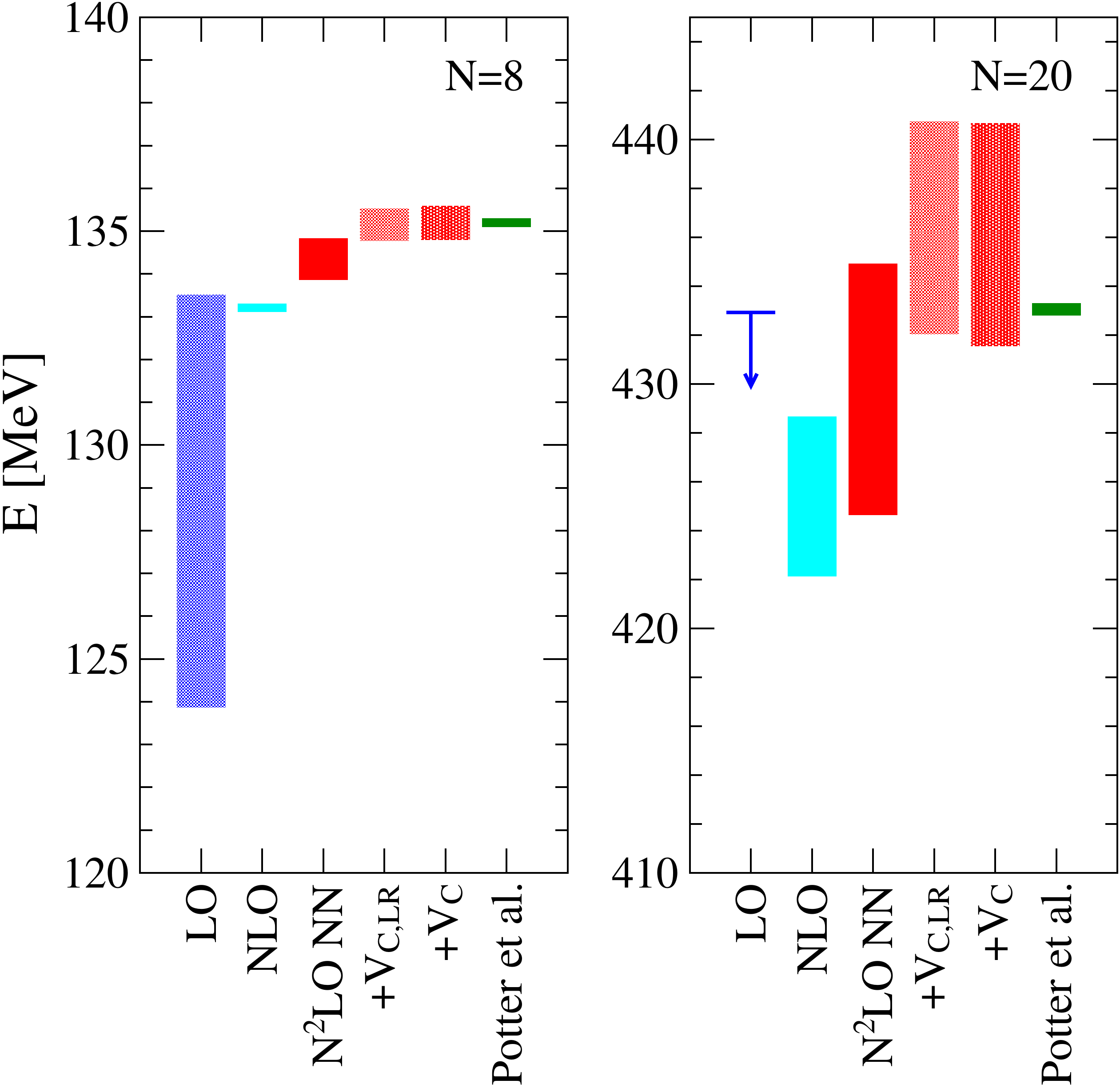} \hspace*{0.2cm}
\includegraphics[height=0.48\textwidth]{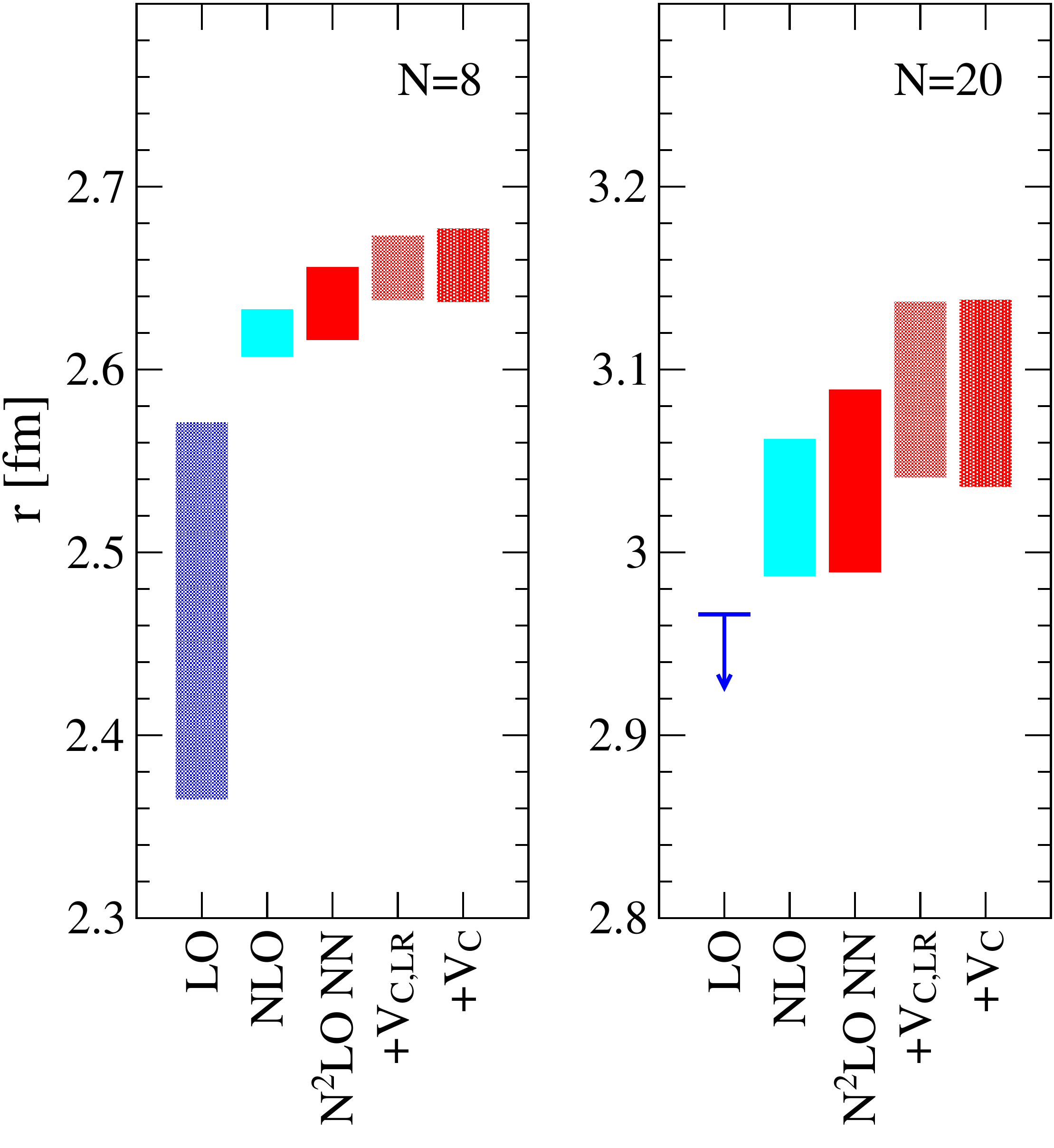}
\caption{Energies and radii of neutron drops with $N=8$ and $20$
neutrons in a harmonic oscillator potential with an oscillator
parameter $\hbar \omega = 10 \mev$ using AFDMC. The same cutoff
is used in the NN and 3N interactions. We give the results at
different orders in the chiral expansion, and at N$^2$LO, for NN
forces only, plus only the LR $c_1$ and $c_3$ parts of $V_C$, and
including also the SR and IR parts of $V_C$. The bands are given by the
cutoff variation $R_0=R_{\text{3N}}=1.0-1.2 \fm$. At LO with the
softer cutoff and $N=20$, the system collapses. We compare our results
with the calculations of Ref.~\cite{Potter:2014}, using
coupled-cluster theory at the $\Lambda$CCSD level, where the band is
given by two different similarity renormalization group (SRG) evolution 
scales (for one initial NN+3N Hamiltonian).\label{fig:drops}}
\end{figure*}

We have investigated the AFDMC energies when choosing different
parameters in the long-range regulator function. In
Fig.~\ref{fig:cutoff_reg_var} we show the dependence of the AFDMC
energy per particle at saturation density as a function of the 3N
cutoff $R_\text{3N}$ on different long-range regulators. Results are
shown for an NN cutoff $R_0=1.2 \fm$. The long-range regulator is
given by $\left(1-e^{-(r/R_{3\text{N}})^{n_1}} \right)^{n_2}$ with
different parameters $n_1$ and $n_2$. We find that the general picture
is independent of the choice of the exponents in the regulator
function. A consistent change of the short-range regulator has only a
negligible effect on the energy.  For different functions, the
position of the plateau varies between $0.8-1.2 \fm$ but the overall
energies at the plateau are comparable, generally ranging between
$12.3-12.5 \mev$.

In Fig.~\ref{fig:n2loEOS} we present the final result of our AFDMC
simulations for the equation of state of neutron matter at N$^2$LO. We
show the energy per particle as a function of density including NN
forces and the 3N $V_C$ interaction. Results are shown for an NN
cutoff $R_0=1.0-1.2 \fm$ and $R_{3\text{N}}$ in the same range. For
the softer NN potential ($R_0=1.2 \fm$, lower lines) we find the
energy per particle to be $12.3-12.5 \mev$ at saturation density for
different 3N cutoffs. The NN-only energy is $11.4 \mev$ and the 3N
$V_C$ has an impact of $\approx 1 \mev$. For the harder NN potential
($R_0=1.0 \fm$, upper lines) we find an energy per particle of
$15.5-15.6 \mev$ compared to $14.1 \mev$ for an NN-only
calculation. Here, the impact of the 3N $V_C$ is $\approx 1.5
\mev$. The variation of the total energy with the 3N cutoff is
$\approx 0.2 \mev$ in our cutoff range and considerably smaller than
the variation with the NN cutoff, because the $R_{\text{3N}}$ range lies
in the plateau described above.

We find the magnitude of the local 3N two-pion-exchange $V_C$ forces
to be at most about $1.5 \mev$ at saturation density, which is is
smaller than a typical contribution of $4 \mev$~\cite{Hebeler:2010a}
in momentum space with nonlocal regulators, including second- and 
third-order corrections. As discussed above, this difference can already be
seen on the HF level and is most likely due to the present local
regulators. This was also observed in the coupled-cluster calculations
of Ref.~\cite{Hagen:2014} where a difference of $2 \mev$ was found for
the neutron-matter energy per particle when choosing local versus
nonlocal regulators with a similar cutoff of $500 \mev$. Following
these findings, local versus nonlocal regulators need to be further
investigated.

In Fig.~\ref{fig:n2lo_comparison} we compare the neutron-matter energy
at N$^2$LO based on the local chiral NN+3N potentials in AFDMC (this
work) with the N$^2$LO calculation of Ref.~\cite{Tews:2013} based on
the EGM N$^2$LO potentials of Ref.~\cite{EGMN2LO} and using many-body
perturbation theory (MBPT), with the particle-particle (pp) ladder
results of Ref.~\cite{Sammarucca:2014} based on the EM N$^2$LO
potential of Ref.~\cite{Marji:2013}, and with results based on the
N$^2$LO$_\text{opt}$ potential of Ref.~\cite{N2LOopt} using
the self-consistent Green's function (SCGF) method~\cite{Carbone:2014}
and using coupled-cluster (CC) theory~\cite{Hagen:2014}. At saturation
density, the AFDMC energies are in general smaller than the other
results, mainly due to the smaller contributions from local 3N
forces. Furthermore, the density dependence of the AFDMC band is
flatter than for the other calculations, which may be explained by
differences in the NN phase shift predictions. We would expect the
results to come closer when including chiral forces at
next-to-next-to-next-to-leading order (N$^3$LO). A comparison of
AFDMC results with MBPT results using the same local potential, as in
Refs.~\cite{Gezerlis:2013ipa, Gezerlis:2014ipa}, will be presented in
a forthcoming paper.

\section{Neutron drops}
\label{sec:5}

Neutron drops in external potentials provide useful constraints for
energy-density functionals and their applications to neutron-rich
nuclei~\cite{Bogner:2011kp,Maris:2013rgq}.
They constitute a simplified model of neutron-rich
nuclei, where the external well simulates the effects 
of the core on the valence neutrons. Their study 
is therefore a natural addition to
homogenous neutron matter.

We have performed AFDMC calculations for the energies and radii of
neutron drops with $N = 8$, $20$, $40$, and $70$ neutrons in a
harmonic oscillator potential with an oscillator parameter $\hbar
\omega = 10 \mev$. For these calculations we used the same cutoff in
the NN and 3N interactions, $R_0=R_{3\text{N}}=1.0-1.2 \fm$.  The
results for the energies and radii are tabulated in
Table~\ref{tab:drops} and shown in Fig.~\ref{fig:drops}. We give the
results at different orders in the chiral expansion, and at N$^2$LO,
for NN forces only, plus only the LR $c_1$ and $c_3$ parts of $V_C$,
and including also the SR and IR parts of $V_C$. The bands in
Fig.~\ref{fig:drops} are given by the cutoff variation
$R_0=R_{\text{3N}}=1.0-1.2 \fm$. We generally find a good
order-by-order convergence of the energies and radii. The band
increases with neutron number; at the level of N$^2$LO+$V_C$ it is
$1\%$ for the energy of $N=8$ neutron drops, $2\%$ for $N=20$, $5\%$
for $N=40$, and $7\%$ for $N=70$. Furthermore, in systems with 
$N>20$ and at low orders, our
calculations do not converge and a collapse occurs (for the
interactions not listed in the table). This is due to the higher 
densities inside the larger neutron drops which are also tabulated for
 the N$^2$LO+$V_C$ Hamiltonian at $r = 0.125$~fm in
Table I. These show that the different particle numbers probe a broad
range of central densities from low densities for $N=8$ to twice
nuclear saturation density for $N=70$, connecting the neutron drop
results with our neutron matter calculations. The higher the (central)
density, the larger the effect of the 3N forces, leading to a collapse
for high densities.

Furthermore, for all $N$ at N$^2$LO the relative contribution of $V_C$
is always larger for the $R_0 = 1.2$~fm potential than for the 1.0~fm
one. For $N=70$ we find $V_C$ to contribute $3.7 \%$ for 1.2~fm versus
$2.7 \%$ for 1.0~fm and for $N=8$ $0.70 \%$ versus $0.56 \%$. This
result is the opposite of what Fig.~6 shows for homogeneous matter:
there the softer NN potential leads to a smaller 3N contribution.
This is due to the higher central densities of the neutron drops for
the softer potentials, leading to larger 3N contributions.

In Fig.~\ref{fig:drops}, we also compare our results at N$^2$LO with
the calculations of Ref.~\cite{Potter:2014} using coupled-cluster
theory at the $\Lambda$CCSD level. The latter results are based on an
SRG-evolved chiral Hamiltonian starting from the N$^3$LO NN potential
of Ref.~\cite{Entem:2003ft} with a cutoff of $500 \mev$ and local
N$^2$LO 3N forces, regulated in momentum space with the same cutoff
value, including also the $V_D$ and $V_E$ parts.  The band is given by
two different SRG evolution scales (for this initial NN+3N
Hamiltonian). We find very good agreement between the two approaches
after inclusion of N$^2$LO 3N forces, whose contribution is small.

\begin{table}[t]
\begin{center}
\caption{Energies (in MeV) and radii (in fm) of neutron drops with $N = 8$, $20$, $40$,
and $70$ neutrons in a harmonic oscillator potential with an
oscillator parameter $\hbar \omega = 10 \mev$. The same cutoff is used
in the NN and 3N interactions. We give the results at different orders
in the chiral expansion, and at N$^2$LO, for NN forces only, plus only
the LR $c_1$ and $c_3$ parts of $V_C$, and including also the SR and
IR parts of $V_C$. For the latter, we also give the
central densities (at $r = 0.125$~fm in fm$^{-3}$). In systems with $N \geqslant 20$ a collapse occurs
at low orders (for the interactions not listed in the table).
\label{tab:drops}}
\begin{tabular}{c|l|c|c|c}
$N$\, & \,Hamiltonian\, & \,$E$\, & \,rms radius & $n_c$\\
\hline
8\, & \,LO(1.0)\, & \,133.51(3)\, & \,2.571(2) &\\
8\, & \,NLO(1.0)\, & \,133.31(3)\, & \,2.633(2) &\\
8\, & \,N$^2$LO(1.0) NN-only\, & \,134.83(3)\, & \,2.656(2) &\\
8\, & \,N$^2$LO(1.0)+$V_{C,\text{LR}}$\, & \,135.53(3)\, & \,2.673(2) &\\
8\, & \,N$^2$LO(1.0)+$V_{C}$\, & \,135.59(5)\, & \,2.677(2) &0.07(1)\\
\hline
8\, & \,LO(1.2)\, & \,123.87(8)\, & \,2.365(2) &\\
8\, & \,NLO(1.2)\, & \,133.11(3)\, & \,2.607(2) &\\
8\, & \,N$^2$LO(1.2) NN-only\, & \,133.86(3)\, & \,2.616(2) &\\
8\, & \,N$^2$LO(1.2)+$V_{C,\text{LR}}$\, & \,134.78(3)\, & \,2.638(2) &\\
8\, & \,N$^2$LO(1.2)+$V_{C}$\, & \,134.80(5)\, & \,2.637(2) & 0.08(1)\\
\hline
20\, & \,LO(1.0)\, & \,432.92(12)\, & \,2.966(2)&\\
20\, & \,NLO(1.0)\, & \,428.67(8)\, & \,3.062(2) &\\
20\, & \,N$^2$LO(1.0) NN-only\, & \,434.93(6)\, & \,3.089(2) &\\
20\, & \,N$^2$LO(1.0)+$V_{C,\text{LR}}$\, & \,440.73(7)\, & \,3.137(2)& \\
20\, & \,N$^2$LO(1.0)+$V_{C}$\, & \,440.67(10)\, & \,3.138(4) & 0.15(1)\\
\hline
20\, & \,NLO(1.2)\, & \,422.13(5)\, & \,2.987(2) &\\
20\, & \,N$^2$LO(1.2) NN-only\, & \,424.64(5)\, & \,2.989(2) &\\
20\, & \,N$^2$LO(1.2)+$V_{C,\text{LR}}$\, & \,432.05(5)\, & \,3.041(2) &\\
20\, & \,N$^2$LO(1.2)+$V_{C}$\, & \,431.55(10)\, & \,3.036(2) & 0.18(1)\\
\hline
40\, & \,NLO(1.0)\, & \,1056.24(26)\, & \,3.459(2) &\\
40\, & \,N$^2$LO(1.0) NN-only\, & \,1071.17(24)\, & \,3.481(2) &\\
40\, & \,N$^2$LO(1.0)+$V_{C,\text{LR}}$\, & \,1094.90(18)\, & \,3.557(2) &\\
40\, & \,N$^2$LO(1.0)+$V_{C}$\, & \,1093.18(32)\, & \,3.556(2) & 0.16(1)\\
\hline
40\, & \,NLO(1.2)\, & \,1018.25(25)\, & \,3.318(2) &\\
40\, & \,N$^2$LO(1.2) NN-only\, & \,1019.44(20)\, & \,3.293(2) &\\
40\, & \,N$^2$LO(1.2)+$V_{C,\text{LR}}$\, & \,1048.70(20)\, & \,3.385(2) &\\
40\, & \,N$^2$LO(1.2)+$V_{C}$\, & \,1045.81(50)\, & \,3.377(2) & 0.20(1)\\
\hline
70\, & \,N$^2$LO(1.0) NN-only\, & \,2235.89(60)\, & \,3.877(2) &\\
70\, & \,N$^2$LO(1.0)+$V_{C,\text{LR}}$\, & \,2302.05(150)\, & \,3.987(2) &\\
70\, & \,N$^2$LO(1.0)+$V_{C}$\, & \,2296.23(111)\, & \,3.991(2) & 0.25(2)\\
\hline
70\, & \,N$^2$LO(1.2) NN-only\, & \,2071.90(70)\, & \,3.593(2) &\\
70\, & \,N$^2$LO(1.2)+$V_{C,\text{LR}}$\, & \,2164.53(180)\, & \,3.730(4) & \\
70\, & \,N$^2$LO(1.2)+$V_{C}$\, & \,2148.75(190)\, & \,3.711(4) & 0.31(2)
\end{tabular}
\end{center}
\end{table}

\section{Summary and outlook}

We have presented local chiral 3N forces at N$^2$LO that are
consistent with the local NN interactions of
Refs.~\cite{Gezerlis:2013ipa,Gezerlis:2014ipa}. We have investigated
the 3N two-pion-exchange contributions to neutron matter both at the
HF level and in AFDMC calculations, including a detailed study of the
regulator dependence. Our results show that present local
regulators lead to less repulsion from 3N forces compared to using the
usual nonlocal regulators. This is already present at the HF level.

In neutron matter, the dependence on the 3N cutoff over the range
$R_{3\text{N}} = 1.0 - 1.2 \fm$ is small compared to the NN cutoff
variation, but for lower 3N cutoffs the system starts to collapse.  We
have also studied the influence of different local 3N regulators and
found that the general picture remains the same.  Our findings lead to
the conclusion that local versus nonlocal regulators have to be
extensively studied. It will be crucial to develop a method of
assessing the quality of local regulators to find improved versions
for these regulators.

We have studied the neutron-matter equation of state for local chiral
NN and 3N interactions and found smaller energies compared to other
calculations, mainly due to less repulsion from 3N forces. We also
simulated neutron drops in an external harmonic oscillator potential
for neutron number $N=8$, $20$, $40$ and $70$ and investigated their
energies and radii for different chiral orders. Our results show
very good agreement with previous coupled-cluster calculations using
chiral potentials~\cite{Potter:2014} (also with local 3N forces).

Work on the determination of the two 3N couplings $c_D$ and $c_E$ for
the local 3N forces is reported in Ref.~\cite{Lynn2015}. The inclusion
of the full N$^2$LO 3N forces will enable novel many-body calculations
of nuclei and nuclear matter with QMC methods based on chiral EFT
interactions.

\begin{acknowledgments}

We thank J.~Carlson, R.~Furnstahl, K.~Hebeler, A.~Lovato, J.~Lynn, and
K.~Schmidt for useful discussions. This work was supported in part by
the ERC Grant No.~307986 STRONGINT, the Natural Sciences and
Engineering Research Council of Canada, the U.S. Department of Energy,
Office of Nuclear Physics, under Contract No. DE-AC02-05CH11231, the
NUCLEI SciDAC program, and the LANL LDRD program. The computations
were performed at the J\"ulich Supercomputing Center. We also used
resources provided by Los Alamos Open Supercomputing and by NERSC,
which is supported by the U.S. Department of Energy, Office of Science,
under Contract No. DE-AC02-05CH11231.
\end{acknowledgments}

\appendix
\section{Coordinate-space expressions}
\label{app:FT}

In momentum space, the N$^2$LO 3N interactions are given by~\cite{3Nforces1,3Nforces2}
\begin{align}
V_C &= \frac{1}{2}\left(\frac{g_A}{2 f_{\pi}}\right)^2 \sum_{\pi(ijk)} \frac{{\fet \sigma}_i\cdot \textbf{q}_i \, {\fet \sigma}_k\cdot \textbf{q}_k}{(q_i^2+m_{\pi}^2)(q_k^2+m_{\pi}^2)} \, F_{ijk}^{\alpha \beta} \, {\tau}_i^{\alpha} \, {\tau}_k^{\beta} \,, \label{eq:Vc} \\
V_D &= -\frac{g_A}{8f_{\pi}^2}\frac{c_D}{f_{\pi}^2 \Lambda_{\chi}} \sum_{\pi(ijk)} \frac{{\fet \sigma}_k\cdot \textbf{q}_k}{q_k^2+m_{\pi}^2} \, {\fet \sigma}_i \cdot \textbf{q}_k \, {\fet \tau}_i \cdot {\fet \tau}_k \,, \label{eq:Vd} \\
V_E &= \frac{c_E}{2 f_{\pi}^4 \Lambda_{\chi}}  \sum_{\pi(ijk)} {\fet \tau}_i \cdot {\fet \tau}_k \,, \label{eq:Ve}
\end{align}
where $\textbf{q}_i=\textbf{p}_i'-\textbf{p}_i$ is the momentum
transfer of particle~$i$ (all other quantities are defined in
Sec.~\ref{sec:2}) and $F_{ijk}^{\alpha \beta}$ includes the different
contributions from the $c_i$'s
\begin{align}
F_{ijk}^{\alpha \beta}&=\delta^{\alpha \beta}\left[-\frac{4c_1m_{\pi}^2}{f_{\pi}^2}+\frac{2c_3}{f_{\pi}^2} \, \textbf{q}_i\cdot \textbf{q}_k \right] \nonumber \\ &\quad +\sum_{\gamma} \frac{c_4}{f_{\pi}^2} \varepsilon^{\alpha \beta \gamma} \, {\tau}_j^{\gamma} \, {\fet \sigma}_j \cdot (\textbf{q}_i \times \textbf{q}_k) \,.
\end{align}
In neutron matter, the 3N contributions simplify because the isospin
structure can be evaluated explicitly, with all ${\fet \tau}_i \cdot
{\fet \tau}_j = 1$ and the $c_4$ part vanishes~\cite{Hebeler:2010a}.

We Fourier transform the 3N interactions with respect to the momentum
transfers of particles $i$ and $k$, which yields the coordinate-space
expression $V^{ijk}$ as a function of $\textbf{r}_{ij}$ and
$\textbf{r}_{kj}$. Because the 3N interactions include a sum over all
permutations, taking a different choice for the momentum transfers
would lead to the same result. However, this will not be the case when
a regulator in momentum space is included before Fourier
transforming. For the $V_E$ contribution this gives
\begin{align}
\quad V_E^{ijk} &= \int \frac{d^3 q_i}{(2\pi)^3} \frac{d^3q_k}{(2\pi)^3} \, e^{i \textbf{q}_i \cdot \textbf{r}_{ij}} \, e^{i \textbf{q}_k \cdot \textbf{r}_{kj}} \, V_E \,, \nonumber \\[1mm]
&= \frac{c_E}{2 f_{\pi}^4 \Lambda_{\chi}} \sum_{\pi(ijk)} {\fet \tau}_i \cdot {\fet \tau}_k \, \delta( \textbf{r}_{ij}) \delta( \textbf{r}_{kj}) \,.
\end{align}

For the Fourier transformation of the $V_D$ contribution one has
\begin{align}
V_D^{ijk} &= \int \frac{d^3 q_i}{(2\pi)^3} \frac{d^3q_k}{(2\pi)^3} \, e^{i \textbf{q}_i \cdot \textbf{r}_{ij}} \, e^{i \textbf{q}_k \cdot \textbf{r}_{kj}} \, V_D \,, \nonumber \\
&= -\frac{c_D g_A}{8f_{\pi}^4 \Lambda_{\chi}} \sum_{\pi(ijk)}  {\fet \tau}_i \cdot {\fet \tau}_k \int \frac{d^3 q_i}{(2\pi)^3} \, e^{i \textbf{q}_i \cdot \textbf{r}_{ij}} \nonumber \\
&\quad \times \int \frac{d^3q_k}{(2\pi)^3} \frac{{\fet \sigma}_k\cdot \textbf{q}_k \, {\fet \sigma}_i \cdot \textbf{q}_k}{q_k^2+m_{\pi}^2} \, e^{i \textbf{q}_k \cdot \textbf{r}_{kj}} \,.
\end{align}
The second integral gives an expression similar to one-pion exchange:
\begin{align}
&\int \frac{d^3q_k}{(2\pi)^3} \frac{{\fet \sigma}_k\cdot \textbf{q}_k \, {\fet \sigma}_i \cdot \textbf{q}_k}{q_k^2+m_{\pi}^2} \, e^{i \textbf{q}_k \cdot \textbf{r}_{kj}} \nonumber \\
=& -\frac{m_{\pi}^2}{12 \pi} X_{ik}(\textbf{r}_{kj}) + \frac{1}{3} \, {\fet \sigma}_i\cdot {\fet \sigma}_k \, \delta(\textbf{r}_{kj}) \,, \label{eq:FT_OPE}
\end{align}
with $X_{ik}(\textbf{r})$ defined in Eq.~(\ref{eq:X}). As a result,
in addition to the one-pion-exchange--contact part, the Fourier
transformation also leads to a contact--contact part in $V_D^{ijk}$:
\begin{align}
V_D^{ijk} &= \frac{c_D g_A}{24f_{\pi}^4 \Lambda_{\chi}}
\sum_{\pi(ijk)} {\fet \tau}_i\cdot {\fet \tau}_k \biggl[ \frac{m_{\pi}^2}{4 \pi} \, \delta(\textbf{r}_{ij}) X_{ik}(\textbf{r}_{kj}) \nonumber \\
&\quad - {\fet \sigma}_i \cdot {\fet \sigma}_k \, \delta(\textbf{r}_{ij}) \delta(\textbf{r}_{kj}) \biggr] \,.
\end{align}

For the $c_1$ part of $V_C$ we have
\begin{align}
V_{C,c_1}^{ijk} &= -\frac{c_1m_{\pi}^2 g_A^2}{2f_{\pi}^4}
\sum_{\pi(ijk)} {\fet \tau}_i \cdot {\fet \tau}_k \int \frac{d^3 q_i}{(2\pi)^3} \frac{{\fet \sigma}_i\cdot \textbf{q}_i}{q_i^2+m_{\pi}^2} \, e^{i \textbf{q}_i \cdot \textbf{r}_{ij}} \nonumber \\
&\quad \times \int\frac{d^3q_k}{(2\pi)^3} \frac{{\fet \sigma}_k\cdot \textbf{q}_k}{q_k^2+m_{\pi}^2} \, e^{i \textbf{q}_k \cdot \textbf{r}_{kj}} \,.
\end{align}
The integrals are readily evaluated using
\begin{align}
&\int \frac{d^3q_i}{(2\pi)^3} \frac{{\fet \sigma}_i\cdot \textbf{q}_i}{q_i^2+m_{\pi}^2} \, e^{i \textbf{q}_i \cdot \textbf{r}_{ij}} \nonumber \\[1mm]
=& -i \, {\sigma}_i^{\alpha} \, \partial^{\alpha} \, \frac{e^{-m_{\pi} r_{ij}}}{4 \pi r_{ij}} = i \, \frac{m_\pi}{4 \pi} \, {\sigma}_i^{\alpha} \, \hat{r}_{ij}^\alpha \, U(r_{ij}) Y(r_{ij}) \,.
\end{align}
This leads to
\begin{align}
V_{C,c_1}^{ijk} &= \frac{c_1 m_{\pi}^4 g_A^2}{2 f_{\pi}^4 (4 \pi)^2}
\sum_{\pi(ijk)} {\fet \tau}_i \cdot {\fet \tau}_k \, {\fet \sigma}_i \cdot \hat{\textbf{r}}_{ij} \, {\fet \sigma}_k \cdot \hat{\textbf{r}}_{kj} \, \nonumber \\
&\quad \times  U(r_{ij}) Y(r_{ij}) U(r_{kj}) Y(r_{kj}) \,.
\end{align}

Next, the Fourier transformation of the $c_3$ part of $V_C$ gives
\begin{align}
V_{C,c_3}^{ijk} &= \frac{c_3 g_A^2}{4 f_{\pi}^4} \sum_{\pi(ijk)}
{\fet \tau}_i \cdot {\fet \tau}_k \int \frac{d^3q_i}{(2\pi)^3}
\frac{{\fet \sigma}_i\cdot \textbf{q}_i}{q_i^2+m_{\pi}^2} \, q_i^{\alpha} \, e^{i \textbf{q}_i \cdot \textbf{r}_{ij}} \nonumber \\
&\quad \times \int \frac{d^3q_k}{(2\pi)^3} \frac{{\fet \sigma}_k\cdot \textbf{q}_k}{q_k^2+m_{\pi}^2} \, q_k^{\alpha} \, e^{i \textbf{q}_k \cdot \textbf{r}_{kj}} \,.
\end{align}
Similar to the Fourier transformation for one-pion exchange in
Eq.~(\ref{eq:FT_OPE}) one gets
\begin{align}
&\int \frac{d^3q_i}{(2\pi)^3} \frac{{\fet \sigma_i} \cdot \textbf{q}_i}{q_i^2+m_{\pi}^2} \, q_i^{\alpha} \, e^{i \textbf{q}_i \cdot \textbf{r}_{ij}} \nonumber \\[1mm]
&= - \frac{m_{\pi}^2}{4 \pi} \, \sigma_i^{\beta} \biggl[\left(\hat{r}_{ij}^\alpha \, \hat{r}_{ij}^\beta - \frac13 \, \delta^{\alpha \beta} \right)T(r_{ij})Y(r_{ij}) \nonumber \\
&\quad + \frac13 \, \delta^{\alpha \beta} \, Y(r_{ij}) - \frac13 \frac{4\pi}{m_{\pi}^2} \, \delta^{\alpha \beta} \, \delta(\textbf{r}_{ij})\biggr] \,.
\end{align}
Combining this leads to
\begin{align}
V_{C,c_3}^{ijk} &= \frac{c_3 g_A^2}{36 f_{\pi}^4} \sum_{\pi(ijk)}
{\fet \tau}_i \cdot {\fet \tau}_k  \\
&\quad \times \biggl[\frac{m_{\pi}^4}{(4 \pi)^2} \, X_{ij}(\textbf{r}_{ij}) X_{kj}(\textbf{r}_{kj}) - \frac{m_{\pi}^2}{4 \pi} \, X_{ik}(\textbf{r}_{ij}) \delta(\textbf{r}_{kj}) \nonumber \\[1mm]
&\quad \quad - \frac{m_{\pi}^2}{4 \pi} \, X_{ik}(\textbf{r}_{kj}) \delta(\textbf{r}_{ij}) + {\fet \sigma}_i \cdot {\fet \sigma}_k  \delta(\textbf{r}_{ij}) \, \delta(\textbf{r}_{kj}) \biggr] \nonumber \,.
\end{align}

Finally, for the $c_4$ part of $V_C$ we obtain
\begin{align}
V_{C,c_4}^{ijk} &= \frac{c_4 g_A^2}{72 f_{\pi}^4} \sum_{\pi(ijk)}
{\fet \tau}_i\cdot({\fet \tau}_k\times{\fet \tau}_j) \nonumber \\
&\quad\times \biggl[ \frac{m_\pi^4}{2i(4\pi)^2} [X_{ij}(\textbf{r}_{ij}),X_{kj}(\textbf{r}_{kj})] \nonumber \\
&\quad - \frac{m_{\pi}^2}{4\pi} {\fet \sigma}_i \cdot ({\fet \sigma}_k\times {\fet \sigma}_j) (1-T(r_{ij})) Y(r_{ij}) \delta(\textbf{r}_{kj}) \nonumber \\
&\quad - \frac{m_{\pi}^2}{4\pi} {\fet \sigma}_i \cdot ({\fet \sigma}_k\times {\fet \sigma}_j) \left(1-T(r_{kj})\right) Y(r_{kj}) \delta(\textbf{r}_{ij})\textcolor{white}{\biggl[} \nonumber \\
&\quad - \frac{3m_{\pi}^2}{4\pi} {\fet \sigma}_i \cdot \hat{\textbf{r}}_{ij} \, \hat{\textbf{r}}_{ij} \cdot ({\fet \sigma}_k \times {\fet \sigma}_j) T(r_{ij}) Y(r_{ij}) \delta(\textbf{r}_{kj}) \nonumber \\
&\quad - \frac{3m_{\pi}^2}{4\pi} {\fet \sigma}_k \cdot \hat{\textbf{r}}_{kj} \, \hat{\textbf{r}}_{kj} \cdot ({\fet \sigma}_j \times {\fet \sigma}_i) T(r_{kj}) Y(r_{kj}) \delta(\textbf{r}_{ij}) \nonumber \\
&\quad + {\fet \sigma}_i \cdot ({\fet \sigma}_k \times {\fet \sigma}_j) \delta(\textbf{r}_{ij}) \delta(\textbf{r}_{kj}) \biggr] \,.
\end{align}

\end{document}